\documentclass[aps,prl,letterpaper,11pt,twoside,tightenlines,nofootinbib,showpacs,preprint,twocolumn]{revtex4}
\usepackage{graphicx}
\usepackage[sort&compress]{natbib}
\usepackage{latexsym}
\usepackage{epsfig}
\usepackage{amsmath}

\newcommand{\be}{\begin{equation}}

\newcommand{\ee}{\end{equation}}
\newcommand{\bea}{\begin{eqnarray}}
\newcommand{\eea}{\end{eqnarray}}
\newcommand{\bef}{\begin{figure}}
\newcommand{\eef}{\end{figure}}
\newcommand{\bce}{\begin{center}}
\newcommand{\ece}{\end{center}}
\newcommand{\E}{\epsilon^{\rm vac}}
\def\lsim{\mathrel{\rlap{\lower4pt\hbox{\hskip1pt$\sim$}}
    \raise1pt\hbox{$<$}}}         
\def\gsim{\mathrel{\rlap{\lower4pt\hbox{\hskip1pt$\sim$}}
    \raise1pt\hbox{$>$}}}         
\begin{document}
\title{Unruh thermalization, gluon condensation  and freeze-out}
\author{P. Castorina$^{1,2}$ and D.Lanteri$^1$}
\affiliation{
\mbox{${}^1$ Dipartimento di Fisica, Universit\`a di Catania, Via Santa Sofia 64,
I-95123 Catania, Italy.}\\
\mbox{${}^2$ INFN, Sezione di Catania, I-95123 Catania, Italy.}\\
}

\date{\today}
\begin{abstract}
The deconfinement transition and the hadronization mechanism at high energy  are related to the quark-antiquark string breaking and the corresponding  temperature  depends on the string tension $\sigma$. In the Unruh scheme of hadron production it turns out  $T = \sqrt{\sigma/2\pi}$, with $\sigma \simeq \E$, the vacuum energy density. In heavy ion collisions at lower energy, i.e. large  baryonchemical potential, $\mu_B$,  the dynamics is dominated by Fermi statistics and baryon repulsion. However one can still consider $\E$ as the relevant physical scale and  its evaluation as a function of the baryon density, in a nuclear matter approach, gives dynamical information on the $\mu_B$ dependence of the hadronization temperature and on the value of the critical end point in the $T-\mu_B$ plane.

\end{abstract}
 \maketitle

\section{Introduction}

The detection of the Hawking-Unruh radiation  has been recently proposed in  various analogue gravity systems \cite{laser1,laser2,laser3,alfredo,liberati,suono}, but the Unruh temperature, $T_U =a/2\pi$ ($a$ is the uniform acceleration),  in those systems is probably still too low \cite{dima}.

On the other hand,  huge accelerations are produced in particle physics and the hadron production in high-energy collisions has been proposed as the Hawking-Unruh  radiation in QCD
\cite{CKS,KT}.

Indeed, it has been shown  \cite{CKS,KT} that the hadron formation occurs at a typical temperature  $T_U = a/2\pi = \sqrt{\sigma/2\pi}/c_0 \simeq 160$ MeV ($\sigma$ is the string tension, $c_0 \simeq 1.1$ \cite{CKS,luscher}) which is universal , i.e. it does not depend on the specific initial setting of the high-energy collisions ($e^+e^-$, hadron-hadron, heavy ions). Moreover its dependence on the quark mass explains \cite{noibec,noi1} the suppression of strange particles production in elementary ($e^+e^-$, hadron-hadron)  collisions with respect to  heavy ion scatterings.

At large energy, the underlying mechanism of the hadron production is the breaking of the quark-antiquark ($q \bar q$) strings with the production of other $q \bar q$ pairs in a self-similar dynamics 
\cite{nussinov,CKS,pajares} and the deconfinement transition is related to the same phenomenon.
In fact in QCD the uniform acceleration of the Unruh effect is due to the Rindler force which describes quark confinement, that is a linear rising potential,$V$, at large distances, $r$, $V \simeq \sigma r$ : for massless quarks with intrinsic transverse momentum $k_T$,  when the
$q \bar q$ distance is such that $\sigma r = 2k_T$ a new pair is produced from the QCD vacuum
\cite{lifetime}. This process gives a thermal spectrum for the final states, with a temperature $T=a/2\pi$ as shown  by many different techniques \cite{unruh1,unruh2,crispino,hosoya,horibe} and in particular by considering  the tunneling through the Rindler event horizon\cite{singleton}, in  analogy with the black-hole calculations \cite{pari}.

The Unruh temperature  $T_U \simeq 160$ MeV refers to high-energy collisions  \cite{CKS} ( center-of-mass energy larger than $\simeq 10-20$ GeV), where the final states are essentially mesons and the hadronization is dominated by the resonance formation and decay.

In heavy ion collisions at lower energy,  the finite baryon density, described by the baryonchemical potential $\mu_B$, has a crucial role and the dynamics is dominated by Fermi statistics and baryon repulsion. In the $T-\mu_B$ plane, the dependence of the hadronization temperature on $\mu_B$ defines the chemical "freeze-out" curve. Lattice QCD simulations, at small density, give the deconfinement critical line and
the relation between the two curves is an interesting problem,  analyzed for example in ref. \cite{fo5,redlich,bbs}. According to previous discussion, they should almost coincide in the small $\mu_B$ region. 

The  chemical freeze-out curve can be described by specific criteria \cite{fo1,fo2,fo3,fo4,fo5,fo6,fo7}. Indeed, a fixed ratio between the entropy density, $s$, and the hadronization temperature, $s/T^3 \simeq 7$, or the average energy per particle, $<E>/N \simeq 1.08$ GeV reproduce the curve in the  
$T-\mu_B$ plane as shown in fig.1, where the percolation model result \cite{fo5}  is also plotted.

\begin{figure}
{{\epsfig{file=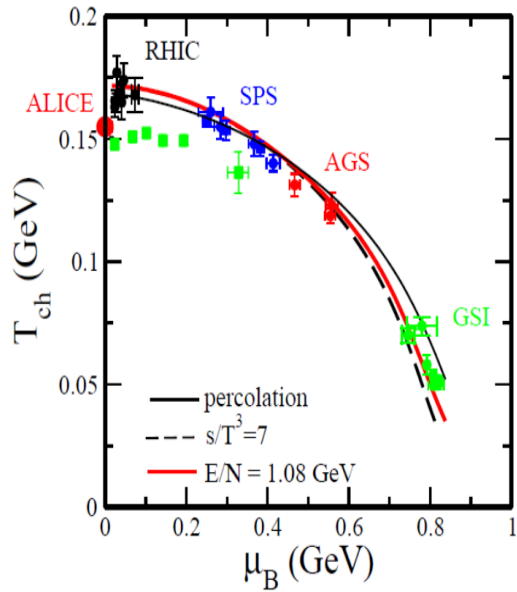,height=8.0 true cm,width=7.0 true cm, angle=0}}
\caption{Freeze-out curve in the statistical hadronization model compared with the criteria discussed in the text. The green squares without error bars are the QCD lattice simulation data.
}}
\end{figure}  

The physical origin of the freeze-out criteria is unclear, but in the Unruh hadronization scheme at $\mu_B \simeq 0$  it is possible to understand them in terms of the breaking of  $q \bar q$ strings 
\cite{CIS}. 

For large baryonchemical potential the interpretation of the freeze-out  and of the QCD critical lines is more difficult and in this regime their evaluation  has been proposed by the generalization of the Unruh hadronization temperature at finite $\mu_B$  in analogy with  black-holes \cite{CKS,grumiller,tawfik} or by the percolation model \cite{fo5}.   

In here, one discusses how the Unruh approach suggests a method to obtain direct dynamical  indications on the $T$ dependence on $\mu_B$, not in analogy with gravity or by general geometric models.

The starting point of our analysis is the relation between the string tension and the vacuum energy density $\E$  \cite{dosch1,dosch2}, essentially the gluon condensate. By considering $\E$  as  the relevant dynamical scale at finite density also, the freeze-out curve originates from  the  decreasing of the gluon condensate by increasing the baryon density  \cite{sou,CBZ,colangelo}. The final results are in agreement with the phenomenological analysis and indicate
$\mu_B \simeq 0.35$ GeV  as  the critical point separating the string dynamics from the high density regime.  However,  $\E$ is not an order parameter for the deconfinement transition and the results have to be considered as an approximation to the freeze-out curve and to the lattice QCD critical $T-\mu_B$ line \cite{lattice1,lattice2}. 

In the next section the freeze-out curve obtained  by analogy with charged black-holes is recalled. Sec.2 is devoted to the relation between the string tension and the gluon condensate and its modification in nuclear matter. Sec.3 contains the results for  the  chemical freeze-out curve and  the  approximations of our analysis  are discussed in  the conclusions.

\section{1. Color confinement and gravitational analogy}

Color confinement  is  a  nonperturbative quantum phenomenon, related to the chromomagnetic properties of the QCD vacuum ( see for example ref.\cite{digiacomo}), producing the squeezing of the chromoelectric field in  quark-antiquark strings  with a constant energy density.

At phenomenological level, confinement is described  by a linear rising potential at large distances, $V = \sigma r$, which corresponds to a constant acceleration, i.e. to a Rindler force.

It is well known that the metric of system in uniform acceleration, the Rindler metric,  is equivalent to the near horizon approximation of the black-holes metric if the acceleration is equal to the surface gravity, $k$. Therefore at local level the correspondence between a  linear rising potential and the dynamics near a black-hole horizon is more than an analogy and 
the idea that an event horizon for color degrees of freedom can be related with  quark confinement is rather natural \cite{recami,salam}. 

Moreover the Hawking radiation is a quantum phenomenon associated with pairs creation near the event horizon and  tunneling, in clear analogy with the string breaking and pair creation in  systems with  uniform acceleration. Indeed, the Hawking temperature is given by $T=k/2\pi$ , equal to the Unruh temperature $T_U =a/2\pi$ for a system in uniform acceleration with $a=k$. 

This point of view suggests  a correspondence at large temperature among quark confinement, Unruh hadronization and black-hole physics. Let us try to apply this correspondence when
a conserved charge, i.e. a chemical potential, is taken into account.

\subsection{Black-holes analogy}

For the Reissner-Nordstrom black hole, with mass $M$ and charge $Q$,  the Unruh-Hawking temperature is given by
\begin{equation}\label{eq:TBH}
 T_{RN}\left(M,Q\right) = T\left(M,0\right)
 \frac{4\,\sqrt{1 - \displaystyle{\frac{Q^2}{G\,M^2}}}}{\left(1 + \sqrt{1 - \displaystyle{\frac{Q^2}{G\,M^2} }}\right)^2 }
\end{equation}
where $T\left(M,0\right)$ is the Hawking-Unruh temperature for a  Schwarzschild black hole.

Also in this case the near horizon approximation corresponds to a Rindler metric with the acceleration equal to the surface gravity and therefore one can evaluate the radiation temperature for an  accelerated observer as a function of the chemical potential.

To obtain the explicit dependence on the chemical potential  it is more useful to start from the first law of black-hole thermodynamics, i.e.
\be
 dM = T_{BH}\,dS_{BH} + \Phi \,dQ + \Omega\,dJ
\ee
where the entropy $S_{BH}$ is defined in terms of the area of the event horizon, $\Phi$ denotes the electrostatic potential and $\Omega$ the rotational velocity
\be
 \Phi = \frac{Q}{R_{RN}} 
 \qquad\qquad
 \Omega = \frac{4\,\pi\,j}{S_{BH}} 
\ee
For a Reissner-Nordstrom black hole ($j=0$),
the Hawking-Unruh temperature (\ref{eq:TBH}) as a function of the electrostatic potential becomes 
\be
 T\left(M,\Phi\right) = T\left(M,0\right)\,
 \left[1 - \left(G\,\Phi^2\right)^2\right]
\ee

The potential $\Phi$ plays the role of a chemical potential and a direct comparison between the Schwarzschild mass-radius relation, i.e. $M=R/2G$, and the Rindler potential energy $V=\sigma r$   suggests the correspondence 
\be
\Phi \rightarrow \mu_B
\ee
and
\be
G \rightarrow 1/2\sigma
\ee
which gives
\be
T \left(\mu, \sigma\right) =
 T\left(\mu=0\right) \,\left[1 - \left(\frac{d\mu^2}{2\,\sigma}\right)^2\right]
\ee
where $d$ is a dimensionless proportionality constant. In Fig.(2) is depicted the behavior of $T(\mu,\sigma)$  
 for $T(\mu=0)=0.155$ GeV and  $d=2\sigma/\bar \mu^2$ corresponding to the value $\bar \mu \simeq 1.1$ GeV for which $T=0$ in the percolation model of ref. \cite{fo5,redlich}.

\begin{figure}
{{\epsfig{file=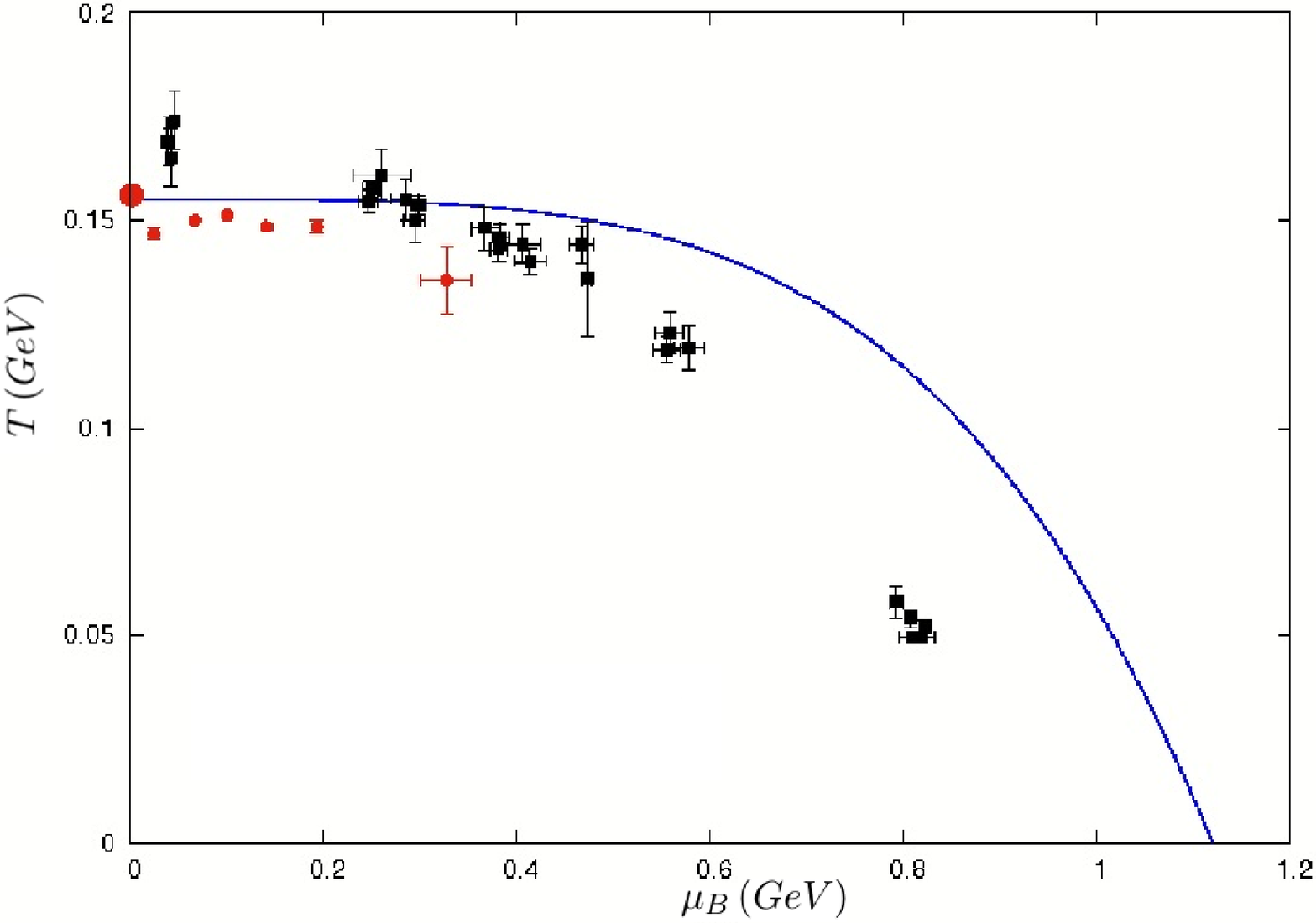,height=8.0 true cm,width=6.8 true cm, angle=0}}
\caption{$T$ as a function of $\mu$ in the black-hole analogy based on eq.(7).
}}
\end{figure}  

The temperature remains rather flat up to large value of $\mu_B$ but a better agreement of the black-hole analogy  with the phenomenological results in Fig.1 can be obtained by assuming
a linear dependence of the string tension on the baryonchemical potential \cite{tawfik} or by considering that near the event horizon the  effective dynamics is 2-dimensional \cite{grumiller}.

\section{2. String tension and gluon condensate at finite density}

The QCD Rindler force is characterized by the string tension which  is the energy density of the vacuum times the transverse string area,
\be
\sigma =  \E\pi r_T^2
\label{s-2}
\ee
with $\E$  essentially due to the QCD gluon condensate,  
\be
\E  \simeq < \frac{\beta(g)}{g} G_{\mu \nu}^a G^{\mu \nu a}>
\ee
where $g$ is the coupling constant and $\beta(g)$ is the QCD $\beta-$function. At $T=0$ and $\mu_B=0$, $\E$ can be evaluated by QCD sume rules \cite{SVZ}: $\E \simeq 0.012 \pm 0.006$ GeV$^4$
\cite{sigma1,sigma2,sigma3,sigma4,sigma5,sigma6}, corresponding to a typical scale $ \simeq {(\E)}^{1/4} = 330$ MeV.
At finite temperature and $\mu_B=0$ the gluon condensate has been studied by lattice simulations \cite{digiacomo2} and it turns out to be $T$ independent  below  the critical temperature 
 and reduced by more than a factor 2 at the transition.

The density dependence of $\E$ has been discussed with different methods \cite{sou,CBZ,colangelo}
and the analysis in  nuclear matter, including nonlinear density effects, gives (see \cite{sou,CBZ} for details)
\be
\begin{split}
\E(\rho,T) =&  \E(0,T) -{8  \rho \over 9} \left(m + \epsilon (\rho,T)\right) -\\ 
& -{8  \rho \over 9} m_s <\bar s s >(\rho,T) 
\end{split}	
\ee
where $\rho$ is the number density, $\E(0,T)$ indicates the gluon condensate at zero density,  $m$ is the nucleon mass,
$\epsilon (\rho,T)$ is the binding energy per nucleon in the medium, $m_s$ is the strange quark current mass, $<\bar s s>$ is the strange quark condensate in the medium,
and the contribution of light quarks has been neglected due to their very small current masses. The strange quark condensate gives a small numerical contribution \cite{sou,CBZ}; it has been included for sake of completeness and its dependence on $T$ and $\mu_B$ will be neglected.   

Since the gluon condensate at $\mu_B=0$ turns out to be  independent on the  temperature and  also $\epsilon(\rho,T)$ is weakly dependent on $T$  up to $T \le 50$ Mev \cite{fiorella} one approximates 
\be
\E(\rho,T) \simeq \E(0) -{8 \over 9} f(\rho) 
\ee
where $\E(0)= \E(0,0)$ and
\be
f(\rho)=\rho(m+\epsilon(\rho)-m_s <\bar s s >)
\ee

The nonlinear effects, contained in $\epsilon(\rho)$,
can be estimated, at finite baryon density and zero temperature, in nuclear matter,
described as a dilute gas of nucleons with a residual
nucleon-nucleon interaction  mainly mediated by mesons. 
In this regime it is a reliable approximation to model nuclear matter as a gas
of nucleons interacting through a static potential. The latter
can be  extracted from the phenomenological analysis
of nucleon-nucleon scattering data. Once the static nucleon-nucleon
interaction is given, the many-body problem can be accurately
solved  and the binding energy as a function of baryon
density can be calculated ( see eg. \cite{fiorella}). 

At high density, the Pauli principle strongly modifies
the two-nucleon scattering process in the medium and  other many-body effects, like the momentum
dependence of the single particle potential and three-body correlations
contribute to the nonlinear terms, which have a certain degree of
extrapolation, since both two-nucleon and three-nucleon forces
must be extended beyond the values of the relative momenta where
they have been phenomenologically checked. However, up to density
3 - 4 times larger than the saturation density the nuclear matter
approximation, obtained along these lines, can be considered
still reliable. We shall use the extrapolation of $\epsilon$ ($\rho$),  evaluated in the previous scheme \cite{fiorella,baldodati}, for symmetric matter and by the  AV18 potential, where one expects the transition from
the string breaking mechanism to the finite density dynamics \cite{fo5,redlich,redlich2}.

The dependence of  $\rho$  and $\epsilon$ on $\mu_B$ 
 in nuclear matter are reported in Fig.(3) and in Fig.(4)
 \cite{fiorella,baldodati} and in both cases there is a linear dependence on $\mu_S=\mu_B-m$ in the range $\mu_S \ge 0.2$ GeV, almost independent on $T$ up to $T \simeq 50$ MeV \cite{fiorella}.
\begin{figure}
{{\epsfig{file=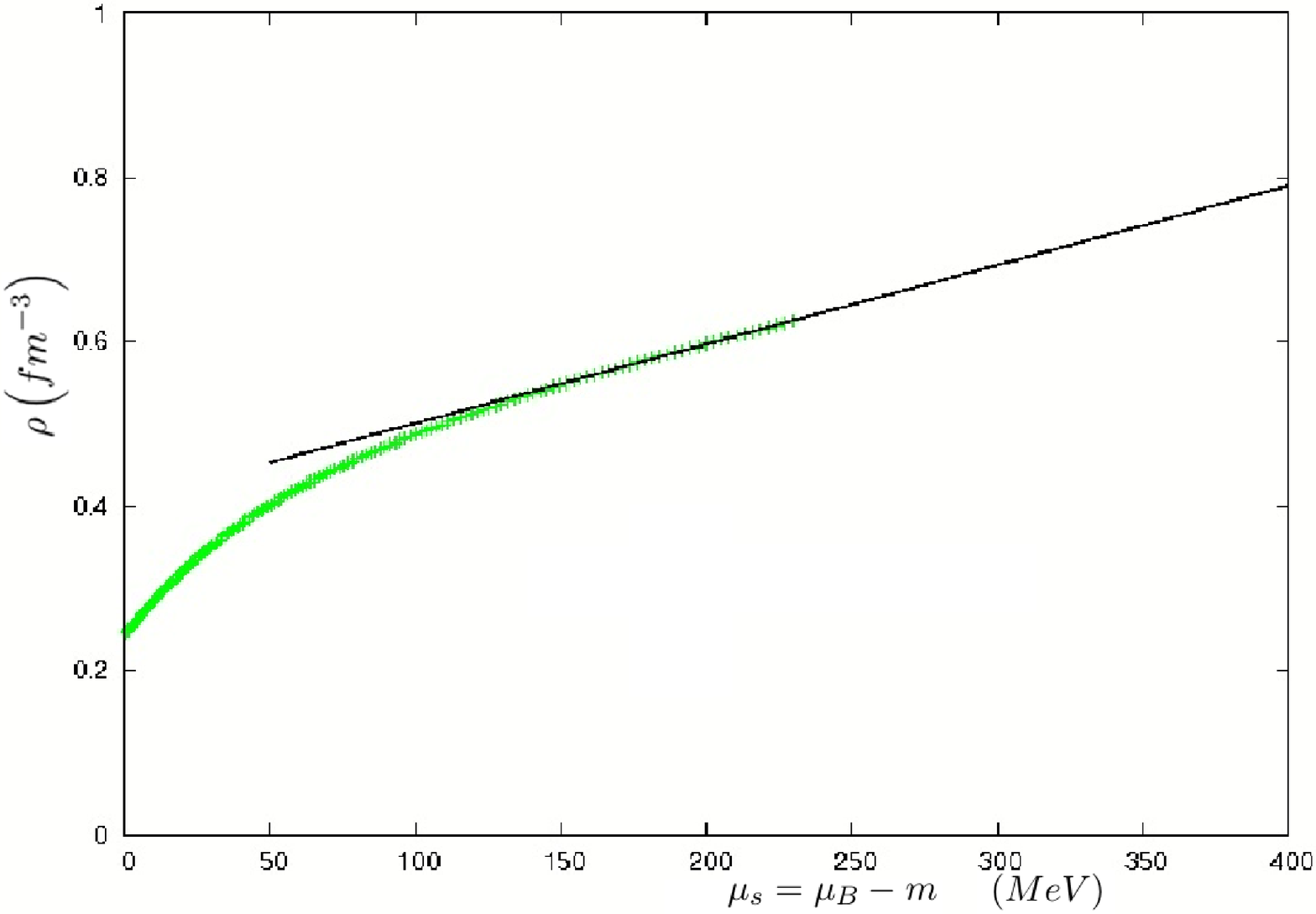,height=8.0 true cm,width=8.0 true cm, angle=0}}
\caption{Extrapolation of the number density as a function of $\mu_S=\mu_B - m$; the green line refers to the nuclear matter results \cite{fiorella,baldodati}.
}}
\end{figure}  

\begin{figure}
{{\epsfig{file=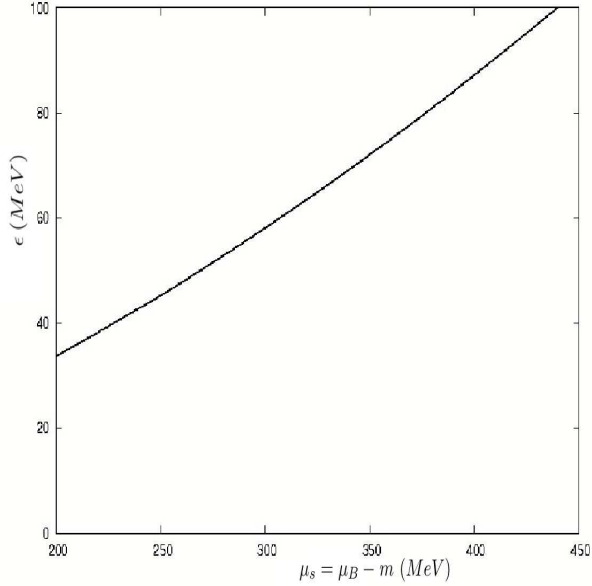,height=8.0 true cm,width=8.0 true cm, angle=0}}
\caption{Extrapolation of the binding density per nucleon as a function of $\mu_S= \mu_B-m$  \cite{fiorella,baldodati}. 
}}
\end{figure}  

The freeze-out curve can be now evaluated on the basis of the previous dynamical analysis.

\section{Evaluating the freeze-out curve}

In the $T-\mu_B$ plane  there are two distinct regimes:  the string breaking and resonance formation at high temperature and  low density and
 the nuclear matter, describing a system of overlapping nucleons with a hard core repulsion,at large $\mu_B$ and small $T$.

\subsection{a. High $T$ , low $\mu_B$ regime}

If one neglects the interrelation of the strangeness and baryon number, at low density 
the resonance production is almost independent of $\mu_B$ \cite{redlich} (taking associated production into account, however, implies with increasing $\mu_B$
an increasing density of strange mesons and thus a slightly decreasing temperature) and then in the range $ \mu_B < \mu_B^0$ one can approximate
\be
T(\mu_B) \simeq T(0)= \sqrt{\sigma/2 \pi}/c_0 \phantom{.........}  \mu_B < \mu_B^0
\ee
where the meaning of $\mu_B^0$ will be clarified later. By eq.(8) and recalling that the string transverse size depends on $\sigma$ according to the relation 
\cite{luscher}
\be
r_T = c_0 \sqrt{2/\pi \sigma}
\ee
one gets
\be
T(\mu) \simeq T(0)= \sqrt{\sigma/2 \pi c_0^2}= \frac{1}{\sqrt{2 \pi c_0}} [2\E(0)]^{1/4}
\ee
 for $\mu_B < \mu_B^0$, which sets $\E(0)$ in terms of $T(0)$.

\subsection{b. Low $T$ , high $\mu_B$ regime}

As the baryon density increases there are more and more nonresonant contributions and the dynamics is described by the  nuclear matter approximation.
The typical string condition in eq.(14) is not valid  anymore; however the vacuum energy density can be considered as the relevant physical scale, which by eq.(8) implies
\be
T(\mu_B) \simeq \sqrt{\sigma} = c [\E(\mu_B)]^{1/2}    \phantom{.........}  \mu_B \ge \mu_B^0
\ee
where $c$ is a constant that can be removed, by continuity, in terms of $T(\mu_B^0)$ to obtain
\be
T(\mu_B)=T(\mu_B^0) [\E(\mu_B)/\E(\mu_B^0)]^{1/2} \phantom{.........} \mu_B \ge \mu_B^0
\ee
with $\E(\mu_B)$ in eq.(11-12), i.e.
\be
T(\mu_B) = T(0) [1 - (8/9)\frac{f(\mu_B)-f(\mu_B^0)}{\E(\mu_B^0)}]^{1/2} 
\ee
The baryonchemical potential for which $T=0$, $\bar \mu$, is solution of the equation
\be
\E(\mu_B^0)= (8/9)[f(\bar \mu)-f(\mu_B^0)]
\ee
i.e. by eqs.(11-12)
\be
\E(0)=(8/9)f(\bar \mu)
\ee
and then $\bar \mu$ does not depend on $\mu_B^0$. 

The previous formulas are reliable at large $\mu_B$ and low $T$ and one needs  a smooth interpolation of the nuclear matter results in this region with the behavior at lower chemical potential and larger temperature. Here one considers $\rho(\mu_B) = a [\mu_B^2-(\mu_B^0)^2]/[\mu_i^2-(\mu_B^0)^2]$, with $a=0.232$ fm$^{-3}$, in the lower density region and the interpolation with nuclear matter results at $\mu_i=900$ MeV. At  low $\mu_B$ the baryon density is very small, i.e. $\rho(\mu_B^0) \simeq 0$, since the system is dominated by mesonic resonances. Moreover $\E(\mu_B^0)$ should have a dependence on the value of $T_0$ and therefore it will be considered as a parameter in eq.(18).

\subsection{c. Freeze-out curve}

In the percolation model the transition from the string breaking and resonance formation to the large $\mu_B$  nuclear matter, describing a system of overlapping nucleons with a hard core repulsion,  is a
first-order mobility or jamming transition \cite{fo5,redlich}. From this point of view the parameter $\mu_B^0$, where the hadronization mechanism changes, can be considered as an estimate of the baryon density at which  the transition between the crossover and the first order phase transition in the deconfinement critical line occurs.

 In Figs.(5) is depicted the result of eqs.(15,18) for $\mu_B^0 =200,400$ MeV, at fixed $T(0)=150$ MeV, which is in agreement with low density, high temperature data but does not fit the
high density data.

By fitting the freeze out curve one gets a good agreement with data, see Fig.(6), for $\E(\mu_B^0)^{1/4} \simeq 190$ Mev and $\mu_B^0 = 353$ MeV  as the crossing point between the two regions in the  $T-\mu_B$ plane.

\begin{figure}
{{\epsfig{file=newfigA.eps,height=5.0 true cm,width=6.0 true cm, angle=0}}
\caption{ Freeze-out curve for $T(0)= 150$ MeV,$\E(\mu_B^0)^{1/4}= 260$ MeV and $\mu_B^0=200,400$ MeV (blue points,black points). The green line gives the constant value for $\mu_B < \mu_B^0$.
}}
\end{figure}

\begin{figure}
{{\epsfig{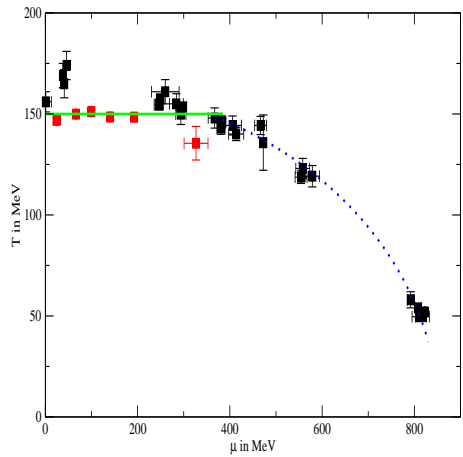}}
\caption{Fit of the freeze-out curve which gives $\mu_B^0 \simeq 0.353$ GeV and $\E(\mu_B^0)^{1/4}= 184$ Mev.
}}
\end{figure}

\section{Conclusions} 

The proposed approach is an attempt to understand the freeze-out curve on  dynamical basis. No black-hole analogy or geometrical models are used.
However, the dynamical analysis can be done at the price of  strong approximations.

The first one concerns the extrapolation of the nuclear matter results from 3-4 times the saturation density,$n_0$, to about 6-8 $n_0$. 
Another simplification is about the dependence of the vacuum energy density on the temperature and on $\mu_B$ , i.e. $\epsilon^{vac}(T,\mu_B)$. We have, indeed,  considered $\epsilon^{vac}$ as a function of $\mu_B$  since the gluon condensate does not depend on $T$, at $\mu_B=0$, below the critical temperature ( lattice data)  and  the density effects in $\epsilon(\rho)$ in eq.(10) are independent on the temperature up to $T \le 50$ MeV. 

With these warnings, the resulting freeze-out curve is reasonable and the evaluation of  the transition point, $\mu_B^0 \simeq 0.35$ GeV, is consistent with previous analyses \cite{fo5,redlich,fo6,cnz}
based on complete different methods.

{\bf Acknowledgements} 
The authors thank M.Baldo and F.Burgio for discussions on the nuclear matter analysis and their more recent results of $\epsilon(\rho)$ and H.Satz for useful comments.


\begin{thebibliography}{99}
\bibitem{lifetime} The corrections to the Unruh temperature due to the finite lifetime of the accelerated observer  have been evaluated in
\cite{MC1,MC2} and they are negligible in the hadronization dynamics \cite{noi1}.
\end{thebibliography}
\end{document}